# Text a Bit Longer or Drive Now? Resuming Driving after Texting in Conditionally Automated Cars

NABIL AL NAHIN CH, University of New Hampshire, USA
JARED FORTIER, University of New Hampshire, USA
CHRISTIAN P. JANSSEN, Utrecht University, Netherlands
ORIT SHAER, Wellesley College, USA
CAITLIN MILLS, University of Minnesota, USA
ANDREW L. KUN, University of New Hampshire, USA

In this study, we focus on different strategies drivers use in terms of interleaving between driving and non-driving related tasks (NDRT) while taking back control from automated driving. We conducted two driving simulator experiments to examine how different cognitive demands of texting, priorities, and takeover time budgets affect drivers' takeover strategies. We also evaluated how different takeover strategies affect takeover performance. We found that the choice of takeover strategy was influenced by the priority and takeover time budget but not by the cognitive demand of the NDRT. The takeover strategy did not have any effect on takeover quality or NDRT engagement but influenced takeover timing.

## 1 INTRODUCTION

In conditionally automated vehicles, drivers will be able to engage in activities that are currently not possible or safe to perform in manually driven vehicles. We expect that drivers will seize this opportunity. Prior studies explored vehicles as a place for work [6, 33, 55–57]. Studies show that in future automated vehicles, people would want to perform various non-driving-related tasks (NDRTs) that require visual and manual resources [51, 63] or tasks that people usually neglect in their daily life [59]. And, as we write this article, there are indications that conditionally automated vehicles (SAE level 3 [8]) will be available in the consumer market [62].

The takeover process in conditionally automated vehicles, that is the transition from automated to manual driving, has been examined in multiple studies (e.g., [25, 41, 42, 53, 58, 64, 65, 74, 78]). In this work, we focus on one important aspect of the takeover process that has not been explored in detail: the strategies that drivers use to switch between driving and NDRTs. Janssen et al. proposed that the process of switching between driving and NDRTs is not a single-step process [28]. Rather, the transfer of control involves a series of steps, similar to an interruption process. This framework allows researchers to study the takeover process in greater detail and examine different strategies drivers use while taking over control of the vehicle. In an empirical test of this framework, Nagaraju et al. found that drivers follow one of two strategies while taking over control of the vehicle [45]. In the *interleaving* strategy, once they are prompted to take back manual control of the vehicle, drivers go back and forth between driving and the NDRT before they finally stop the NDRT and start driving. In the *suspension* strategy drivers stop the NDRT and immediately switch to the driving task.

Reading and writing text messages or emails are among the most common NDRTs that drivers currently perform while driving [63]. It is also one of the most common NDRTs that drivers are interested in performing in future automated vehicles [51, 63]. However, we do not fully understand how different aspects of these NDRTs and various elements of the driving context can influence drivers' strategies while transitioning from the NDRT to driving. We also do not know how different takeover strategies (*interleaving* and *suspension*) affect takeover and NDRT performance.

To find answers to these questions, we conducted two driving simulator studies. In these studies, participants performed different texting-related NDRTs while the car was in automated driving mode and switched to manual driving when takeover requests were presented. We examined how different types of texting conversations, the priorities assigned to the NDRT and driving, and the allowed takeover time influence takeover strategy regarding interleaving between driving and texting. We also analyzed how takeover strategy affects takeover performance, as well as engagement in texting conversations. The contribution of this study is extending existing knowledge on the takeover process in conditionally automated vehicles.

## 2 RELATED WORK

### 2.1 Multitasking and Task Interleaving

People often have to "multitask" and switch between various activities while working [9], sometimes as frequently as once every 2-3 minutes on average [20]. While multitasking, people interleave between multiple tasks to improve overall performance and maximize the marginal rate of return [11].

Over the years, numerous studies have explored how factors like difficulty and priority of tasks influence people's strategies for interleaving and how that information can be used to predict when someone might decide to switch tasks. For example, Duggan et al. suggested that the perceived marginal rate of return may be used to determine when people interleave between tasks [11]. People can estimate the marginal rate of return based on how easy or important the task is or how close they are to completing the task. The strategic task overload management (STOM) model described by Wickens et al. predicts the decision to switch tasks during sequential multitasking based on similar task attributes; difficulty, priority, interest, and salience [71]. Other studies also found similar effects of cognitive demand or task difficulty [21, 31, 32], and potential reward or task priority [21, 50, 52] on people's decision of choosing which task to perform and when to switch between tasks.

### 2.2 Texting in Cars

People's frequent engagement in various activities while driving and their desire to continue to do so in highly automated vehicles has been well documented in prior studies [51, 59, 63]. For some NDRTs, like reading or typing emails or text messages, people often use hand-held devices like smartphones. Texting while driving is a common occurrence among adult drivers around the world [16, 46], and it is especially frequent among young drivers [47, 48]. This is a serious safety concern, as thousands of fatalities are estimated to have resulted from texting while driving [73].

Texting and using a smartphone while driving has been shown to adversely affect various measures associated with safe driving. A meta-analysis of 28 experimental studies found negative effects of texting while driving on drivers' reaction time, gaze behavior, lateral and longitudinal control of the vehicle, ability to detect traffic events, and risk of collisions [5]. Use of hand-held devices like smartphones while driving has been linked to increased cognitive demand [1] and lack of spatial awareness [60]. This, in turn, results in slower reaction time [1, 36], reduced steering wheel control [3], and ultimately higher probability of collisions [24, 54, 67].

Similar adverse effects of texting have been observed on drivers' ability to take back control of the vehicle from automated driving. Tasks requiring visual attention or smartphone interaction can lengthen the time drivers need to stop the NDRT and resume driving [10, 13, 76]. In contrast to these findings, in their study, Zeeb et al. found that similar NDRT did not affect takeover time [77]. However, it did affect takeover quality in terms of drivers' ability to maintain lane position after taking back control of the vehicle.

These adverse effects of such NDRTs on driving performance have often been attributed to the physical requirements of the tasks, like the manual manipulation of devices. In a meta-analysis of 129 studies focused on conditionally automated vehicles, Zhang et al. found a strong influence of hand-held devices on takeover time, whereas the effect of hands-free NDRTs was small [78]. However, other studies show that hands-free NDRTs can have similar negative effects on driving [54, 60], which suggests that cognitive attributes of NDRTs may have a stronger influence. By examining 10 realistic NDRTs, Lee et al. found that the cognitive load of NDRTs can adversely affect lateral and longitudinal control of the vehicle during takeover [37]. In contrast, the physical and visual attributes of NDRTs did not have a significant effect [37]. Similarly, Kaye et al. found no significant difference between the influence of hand-held device manipulation and hands-free working memory task [30]. Cognitive demand for different NDRTs is not the same, and retrieving information from memory can interfere more in complex driving situations [24].

Another important factor to consider while examining the takeover process is the amount of time available to the driver to take back control of the car. The time budget can influence both the timing and quality of the takeover. With a shorter time budget, drivers tend to make decisions quicker and react faster. But it hinders their ability to scan the surroundings and increases the risk of collisions [17]. With a more extended time budget, drivers take longer to take back control of the vehicle [68, 78], but their takeover performance improves [18].

In addition to driving and NDRT performance, several studies also investigated different strategies people use for multitasking in a manually-controlled vehicle. Similar to task switching in other domains, people's strategy for switching between driving and NDRTs depends on the difficulty of NDRT, the complexity of the driving scenario, and the priority or potential rewards of driving and NDRTs [4, 23, 26, 27, 61]. Drivers also decide at what stage to suspend NDRT and focus on driving based on performance objectives of driving and NDRT [4, 26, 27].

## 3  RESEARCH QUESTIONS

These studies demonstrate that multitasking and task interleaving have been examined extensively in different driving scenarios but not during takeovers. Research on takeovers in conditionally automated vehicles has been mostly focused on takeover performance, not takeover strategy, in terms of interleaving between driving and NDRTs. Since people often interleave while taking over control of the vehicle from automated driving [28, 45] and interleaving can affect both driving and NDRT performance, we need to investigate interleaving during takeovers. Among other factors, cognitive demands, priority, the available time for a takeover, and stage of NDRT can influence multitasking strategy and takeover in conditionally automated vehicles.

We explore the effect of these factors in two driving simulator experiments. In the first experiment, we focus on texting conversations as NDRT, and examine the following research questions:

RQ1.1  How do different cognitive demands of texting conversations influence takeover strategy?
RQ1.2  How do different takeover time budgets influence the strategy for taking over from texting conversation?
RQ1.3  How do different takeover strategies affect drivers' performance of takeover from texting conversation?
RQ1.4  How do different takeover strategies affect drivers' engagement in texting conversations?

In the second experiment, we focused on a multi-step texting task and we examine the following research questions:

RQ2.1  How do the different priorities influence the takeover strategy from a multi-step texting task?
RQ2.2  How do the different takeover time budgets influence the takeover strategy from a multi-step texting task?
RQ2.3  How do different takeover strategies affect drivers' engagement in a multi-step texting task?
RQ2.4  How do different takeover strategies affect drivers' performance of takeover from a multi-step texting task?
RQ2.5  How does the NDRT stage at TOR influence the driver's decision of when to stop NDRT?

## 4 EXPERIMENT 1: COGNITIVE DEMAND OF NDRT

We conducted a within-subjects driving simulator experiment to examine drivers' takeover strategy while texting. We investigated whether their takeover strategy choice depended on the cognitive demand of texting conversations (RQ1.1) and the amount of time available for the takeover (RQ1.2). We also evaluated the effect of takeover strategies on takeover performance (RQ1.3) and on engagement in texting (RQ1.4).

### 4.1 Participants

We recruited 24 people ($Man = 15, Woman = 9$) with an average age of 23.71 years ($SD = 4.34$) to participate in this experiment. All the participants reported that they held valid driver's licenses. Each participant received a $20 gift card as compensation for their time. The study was approved by the Institutional Review Board (IRB) of the University of New Hampshire.

### 4.2 Apparatus and software

Participants operated a driving simulator and used a smartphone for the NDRT. They also wore an eye tracker for the whole duration of the experiment. We used Ergoneers D-lab to synchronously collect data at 60 Hz from the driving simulator and the eye-tracker. Figure 1 shows the experiment setup for the experiments, and the smartphone interface for the NDRT is shown in Figure 2.

*4.2.1 Driving simulator.* We used the miniSim driving simulator developed by the University of Iowa Driving Safety Institute (DSRI). Participants were positioned in the driver's seat of the simulator fit with Fanatec steering wheel, pedals, B-Box motion system, and a usable instrument panel for actions such as using blinkers, adjusting mirrors, and toggling the activation of automated driving mode. The simulator displays the driving view on three 48-inch screen monitors, and a fourth 18.5-inch monitor is positioned right behind the steering wheel to display the instrument panel. We collected driving data at a 60 Hz rate. We also recorded the time when takeover requests were presented or when automated and manual driving started.

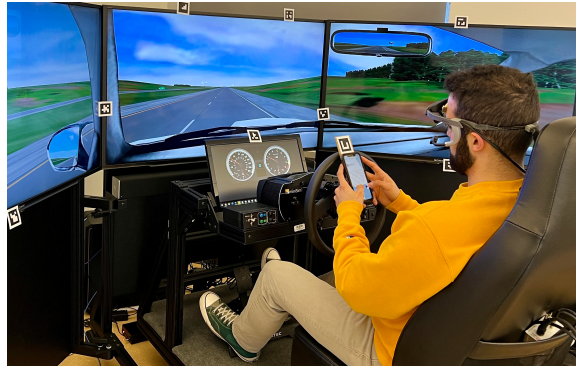

Fig. 1. Participant is wearing the eye tracker and performing NDRTs on a smartphone while the vehicle is in automated driving mode.

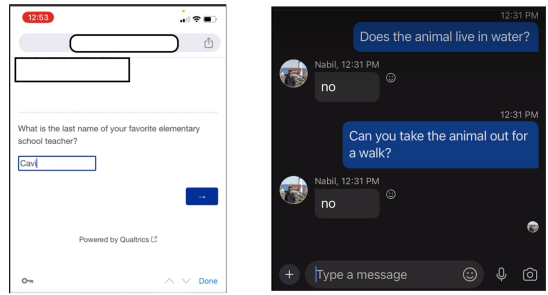

Fig. 2. Participants engaging in NDRTs in the first (left) and second (right) experiment.

*4.2.2 Eye tracker.* We used an Ergoneers Dikablis head-worn eye tracker to collect gaze positions at a 60 Hz rate. To determine when participants were looking at the NDRT or driving task, we defined two areas of interest (AOI) captured by placing 2-D markers on different places of the driving simulator and attached to the top of the smartphone. The AOI for the driving task includes the three driving scene monitors, steering wheel, and instrument panel monitor. The AOI for the NDRT includes looking at the smartphone. Since the participants performed the texting task on a hand-held smartphone, the NDRT AOI moved with the marker attached.

*4.2.3 Smartphone.* Participants were given a 6.1-inch smartphone to complete the NDRT. We changed the smartphone's settings to ensure the display does not go to sleep after any period of inactivity. Smartphone activities were also recorded to analyze participants' engagement in the NDRT.

### 4.3 Task

*4.3.1 Driving task.* In each drive, participants drove a simulated vehicle in a simple driving scenario for approximately eight minutes. The simulated drive was on a straight two-lane highway in daylight, with no traffic other than a lead vehicle programmed to maintain a constant speed of 104.6 km/h (65 mph). Each lane was 3.66 meters (12 feet) wide. Participants were asked to maintain a safe following distance. Participants drove routes of 60-second segments before the vehicle took over driving responsibilities with automated driving. The simulator would alert the driver of this switch with an audible beep followed by the voice alert message "Automated Driving Engaged."

*4.3.2 Non-driving-related task.* In this experiment, we focused on texting since it is a common task people do in cars that can affect driving and takeover performance. Iqbal et al. categorized phone conversations into three categories based on their cognitive demand: *assimilation, retrieval, generation* [24]. We followed their approach in the design of our texting-based NDRTs. Thus, in our experiment, during the *assimilation* task, participants were asked to read short news articles with a mean word count of 119 (e.g., "This January, temperatures across Europe reached an all-time high."). This task required participants to take in new information. They were then tasked with answering two short comprehension questions about the article. The *retrieval* task asked participants to answer simple personal questions that required them to retrieve some information from their memory (e.g., "What was the last name of your first boss?)". The *generation* task required them to generate new information, like driving directions between two points of interest (e.g., "Please give directions from your home to your favorite local restaurant."). These points of interest were chosen to be locations that one would commonly visit or know the location of; the university library, local grocery store, and doctor's office. These tasks were designed to elicit different types of cognitive demands. These questions were presented one at a time on the smartphone. We used Qualtrics, a web-based survey tool to design these tasks. Participants were able to scroll and type to complete these tasks, similar to any other texting application.

*4.3.3 Switching between tasks.* Participants were given 60 seconds to perform the NDRT while engaged in automatic driving. At the end of the 60 seconds, the system would issue an alert message takeover request to the participant. Participants could take over driving by pressing a steering wheel button, brake, or acceleration pedal. In long takeover conditions of 30 seconds, they would first hear a beep and a voice saying, "Take over the vehicle within the next 30 seconds". If the driver did not take back control after 20 seconds, the system would issue a final alert with a beep that said, "Take over control of the vehicle", which indicated the driver must take back control in 10 seconds or less. In the short takeover condition of 10 seconds, the driver would only be issued the final alert for a takeover request. The system disengaged automated driving if the driver did not take control once the maximum allowed takeover time had passed (30 seconds or 10 seconds, depending on the condition). When the system switched to manual driving, an audio message was played: "Manual Driving Started."

## 4.4 Procedure

To begin the study, participants read and signed an electronic consent form and completed a demographic survey. Then, participants were given an information sheet explaining the steps of the experiment. The experimenter answered any questions about the procedure the participants might have asked. Next, we trained participants on NDRT and the operation of the driving simulator. This allowed the participant to familiarize themselves with driving in the simulator and the process of engaging in the NDRT. Additionally, participants practiced taking back control of the vehicle from automated driving. The entire training took approximately 10 minutes.

Once trained, the participants completed six drives under different conditions (text type (3) x available time (2)) using the same driving scenario. The order of the conditions presented was counterbalanced individually for text type and available time across participants. In each drive, participants switched back from the NDRT to driving three times. The eye tracker was calibrated and validated before each drive. Following each drive, the experimenter asked the participants if they were feeling any discomfort and whether they wanted to take a break before starting the next drive.

Participants could keep the smartphone on a stool on their right or left, depending on their preference. While texting, participants were asked to hold the smartphone in front of the steering wheel. The experimenter maintained a checklist to ensure that the same steps were followed for each participant and the participants followed the instructions.

### 4.5 Measures

*4.5.1 Takeover strategy.* We identified two strategies: *interleaving* and *suspension*. In *interleaving* strategy, drivers looked at the driving scene after the takeover request and returned their gaze to the smartphone before finally taking back control of the car from automated driving. In *suspension* strategy, drivers stopped NDRT after the takeover request and resumed driving without going back and forth between texting and driving.

*4.5.2 Takeover performance.* Takeover performance was evaluated in numerous prior experiments using various measures related to the timing and quality of the takeover [40, 70]. We evaluated takeover timing using automation deactivation time (ADT) and gaze reaction time (GRT). These measures have been used in similar experiments to examine reaction times related to takeovers [10, 12, 17]. Automation deactivation time measures how long drivers take to disengage automated driving after the initiation of the takeover request. Gaze reaction time measures how long drivers took to look at the driving-related AOI after the takeover request.

The quality of takeovers was assessed using two measures; the standard deviation of velocity (SDV) and the standard deviation of lateral position (SDLP). These are standard measures used in prior experiments to evaluate vehicle control during takeovers [7, 42, 43, 69, 72]. Prior studies show that SDLP varies between subjects for on-the-road tests, but it is a stable and reliable measure within subjects [29, 66], even for PC-based driving simulators [39]. After taking over control of the vehicle from automated driving, it can take the driver up to 40 seconds to stabilize control [42]. So we calculated SDLP and SDV for the 30-second period immediately after the disengagement of automated driving.

*4.5.3 NDRT engagement.* The engagement in NDRT was calculated as the number of questions participants attempted in each automated driving phase. The automated driving phase was defined as the time between the start and end of automated driving. This included the time between the presentation of the takeover request and the time when the participant took over control.

### 4.6 Data analysis and results

Each of the 24 participants switched from NDRT to driving 18 times (text type (3) x available time (2) x takeovers per condition (3)). Thus, we expected to process 24 x 18 = 432 data points (takeovers) for each research question. However, due to technical problems, we could not record driving-related data for a total of 20 drives and eye-tracking data for two drives. We also discarded data from 14 takeovers for which the eye-tracking data were inaccurate. Thus, we analyzed data from 24 participants and 412 takeovers for the takeover strategy (only eye-tracking data was used to determine the takeover strategy). Also, we analyzed data from 22 participants and 356 takeovers for takeover performance (both eye-tracking and driving-related data were needed).

We used a mixed-effects logistic regression approach to analyze whether the probability of adopting a particular takeover strategy (interleaving, suspension) depends on different cognitive demands of texting conversation and the available time for the takeover. We regressed the takeover strategy on the takeover time budget and cognitive demand. We used a mixed-effects linear regression approach to examine the relationship between the takeover strategies and the takeover performance (SDLP, SDV, automation deactivation time, gaze reaction time) and engagement in texting (number of texts). For these models, the type of texting task and available time were treated as covariates and included as fixed effects to control for their effect on dependent variables. Participant ID was included as a random effect in all the models to account for baseline differences among individuals. We used the R package lme4 [2] for the models, and the significance was assessed in the form of chi-square tests using R package car [14]. We also estimated marginal means using the emmeans package [38].

Table 1. Effects of texting conversations and time budget on the probability of adopting the interleaving strategy; estimates are in logit.

| Predictors | | Interleaving probability $\beta$(SE) | $\chi2(p)$ |
|---|---|---|---|
| | Intercept | -3.82(0.71) | |
| Tasks | Generation | 0.29(0.77) | 0.55(0.76) |
| | Retrieval | 0.55(0.75) | |
| Time budget | 30 sec | 1.70(0.68) | **6.20(0.01)** |
| Tasks:Time budget | Generation:30 sec | -0.24(0.93) | 0.18(0.91) |
| | Retrieval:30 sec | -0.38(0.90) | |

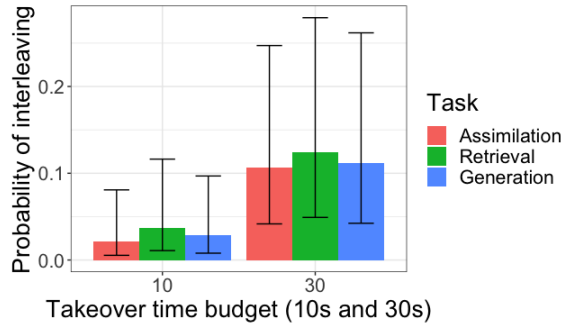

Fig. 3. Relationship between takeover time budget and probability of adopting interleaving strategy during takeovers. Error bars present 95 Percent Confidence Limits.

*4.6.1 Effects of different texting conversations & takeover time-budget on takeover strategy (RQ1.1 & RQ1.2).* We found that the probability of drivers adopting the interleaving strategy is significantly affected by the takeover time budget but not by the different cognitive demands of texting conversations (Table 1). Drivers were more likely to go back and forth between driving and texting during the takeover when the allowed time for the takeover was 30 seconds ($M = 0.11$) compared to when it was 10 seconds ($M = 0.03$) (Figure 3). However, the probability of adopting an interleaving takeover strategy remained similar for assimilation, retrieval, and generation texting conversations. There was also no interaction effect between task type and time budget.

*4.6.2 Effects of takeover strategy on takeover performance & engagement in texting conversations (RQ1.3 & RQ1.4).* Our analysis shows no significant relationship between takeover strategy and driving performance after the takeover (SDLP: $\chi2(1) = 4.71, p = 0.49$, SDV: $\chi2(1) = 3.69, p = 0.055$). Control variables (type of texting task, takeover time budget) did not have a significant effect either. However, takeover strategies influenced takeover timing in terms of both ADT and GRT (ADT $\chi2(1) = 22.23, p < 0.001$, GRT $\chi2(1) = 46.99, p < 0.001$). While adopting an interleaving strategy, drivers took longer to take over control of the vehicle but were quicker to glance at the driving scene after the takeover request

Table 2. Marginal means estimated from the corresponding models. Results are averaged over the three levels of tasks and two levels of takeover time budget.

|  |  | Suspension Mean(SE) | Interleaving Mean(SE) |  |
| --- | --- | --- | --- | --- |
| Takeover quality | SDLP (cm) | 20.40(1.40) | 19.50(1.74) | t(344)=0.69, p=0.49 |
|  | SDV (kmph) | 3.27(0.34) | 3.92(0.45) | t(347)=-1.92, p=0.06 |
| Takeover timing | ADT (sec) | 8.04(0.85) | 11.87(1.10) | **t(347)=-4.70, p<0.001** |
|  | GRT (sec) | 7.26(0.77) | 1.36(1.07) | **t(350)=6.83, p<0.001** |
| NDRT engagement | Questions attempted | 3.18(0.18) | 3.5(0.28) | t(351)=-1.26, p=0.21 |

(Table 2). Among the control variables, only the time budget was a significant predictor of takeover timings (ADT: $\chi2(1) = 216.47, p < 0.001$, GRT: $\chi2(1) = 125.45, p < 0.001$). We did not find any relationship between takeover strategy and NDRT engagement ($\chi2(1) = 1.61, p = 0.20$). However, the control variables were significant predictors of NDRT engagement (text type: $\chi2(1) = 627.54, p < 0.001$, time budget: $\chi2(1) = 4.95, p = 0.02$).

## 5 EXPERIMENT 2: PRIORITY OF NDRT

Previous work has shown that interleaving patterns might also differ if the task has a hierarchical structure and if drivers have different priorities [26, 27]. As the task used in experiment 1 did not have such a hierarchy, we here test performance on a multi-step texting task. We conducted a within-subjects driving simulator experiment to examine whether drivers' takeover strategy while performing a multi-step texting task depends on the priority (RQ2.1) and whether the available time for takeover moderates the effect of priority (RQ2.2). We also investigated whether the takeover (RQ2.4) and NDRT (RQ2.3) performance depended on the strategy they chose and the influence of NDRT stage on takeover decision (RQ2.5). Similar to the first experiment, participants of the study operated a vehicle using a driving simulator and switched between automated and manual driving. During automated driving, participants performed an NDRT, and they took back control of the vehicle when takeover requests were presented. Participants were instructed to prioritize either driving or NDRT for each drive, and the takeover time budget was either 10 or 30 seconds. Priority was manipulated through verbal instructions without using any additional incentive. Each participant completed the same driving scenario four times for these four (2x2) conditions. The order of the conditions was counterbalanced.

### 5.1 Participants

We recruited 24 participants ($Man = 17, Woman = 7$) with an average age of 25.67 years ($SD = 4.98$) for this experiment. All the participants held valid driver's licenses and received a $20 gift card for their participation. The study was approved by the Institutional Review Board (IRB) of the University of New Hampshire.

### 5.2 Apparatus and software

For this experiment, we used the same apparatus and software we used in the first experiment (described in section 4.2). The experiment setup and interface for performing NDRT are shown in Figure 1 and 2, respectively.

### 5.3 Task

In the second experiment, the driving task and the process of switching between driving and NDRT were the same as in the first experiment (described in section 4.3). For the NDRT in this experiment, we focused on a multi-step texting task called the twenty-question task (TQT) [44]. This task has been widely used in similar studies as an NDRT [19, 22, 34, 35, 41, 49]. The objective of this task is to guess an item by asking as few yes-no questions as possible. This task requires participants to solve problems by planning, generating information, and using working memory, similar to many everyday tasks people perform. Participants typed their responses on Skype to perform the task. If the participants were not able to guess the item in the time allocated for the automated driving phase and takeover, they could resume the task from where they left off when the next automated driving phase started. Participants were allowed to give up on guessing an item and move to the next item. The priority (driving, NDRT) and takeover time budget (10s, 30s) were varied in four drives. Participants guessed an item from 10 items in either the "fruits and vegetables" or "animals" categories. We used the most generic form of any item instead of a particular type (e.g., cat instead of a particular breed of cat). Participants did not have access to the list while driving or performing the NDRT. In driving-focused and NDRT-focused conditions, the participants were instructed to prioritize driving and TQT, respectively.

### 5.4 Procedure

The procedure was the same as the procedure of the first experiment (section 4.4). The only difference was that the participants completed four drives in different conditions in this experiment instead of six drives in the first experiment.

### 5.5 Measures

The takeover strategy and performance were evaluated using the same measures as described in section 4.5 for the first experiment. For the NDRT engagement, we evaluated how many questions participants asked and how many items they correctly guessed in each automated driving phase. The stage of the NDRT was calculated based on how many questions the participant asked to guess the current item at the time when TOR was presented or when the driver took over driving. For example, if the participant had asked two questions to guess the current item when the TOR was presented, the stage of the NDRT would be considered two. The stage was considered to be 0 if the driver guessed an item (correctly or incorrectly) and did not start asking questions to guess the next item.

### 5.6 Data analysis and results

In this experiment, each participant switched from NDRT to driving 12 times (priority (2) x available time (2) x takeovers per condition (3)). Thus, we expected to analyze a total of 24 x 12 = 288 data points (takeovers). However, due to technical issues, we could not collect driving-related data from three participants and eye-tracker data for four drives. We discarded data for 17 takeovers for which either the participant did not follow the instructions (changed lane) or the eye-tracker data were inaccurate. Thus, we analyzed data from 24 participants and 259 takeovers for the takeover strategy (only eye-tracking data were needed to determine the takeover strategy) and data from 21 participants and 227 takeovers for takeover performance (both eye-tracking and driving-related data were needed).

We used a mixed-effects logistic regression approach to evaluate the effect of priority (NDRT or driving) and takeover time budget on the probability of adopting different takeover strategies. We included an interaction term between priority and time budget to examine whether the effect of priority is moderated by the time allowed for takeovers. To investigate the effect of takeover strategies on takeover performance and NDRT engagement, we used a mixed-effects linear regression approach. To control for the effects of priority and time budget on dependent variables, we included them as fixed effects in the models. Similar to the first experiment (see section 4.6), participant ID was included as a random effect in all the models, and the same R packages were used for the analysis. Lastly, we used the test of proportions to evaluate whether drivers' decision to complete the current NDRT (stopping at stage 0 of NDRT) and then resume driving is influenced by the stage of the NDRT at the time of TOR.

Table 3. Effects of priority and time budget on the probability of adopting the interleaving strategy; estimates are in logit.

| Predictors | | Interleaving probability | |
|---|---|---|---|
| | | $\beta(SE)$ | $\chi 2(p)$ |
| | Intercept | -2.99(0.58) | |
| Priority | NDRT | 1.63(0.61) | **7.26(<0.01)** |
| Time budget | 30 sec | 2.83(0.61) | **21.18(<0.001)** |
| Priority:Time budget | NDRT:30 sec | -1.71(0.74) | **5.30(0.02)** |

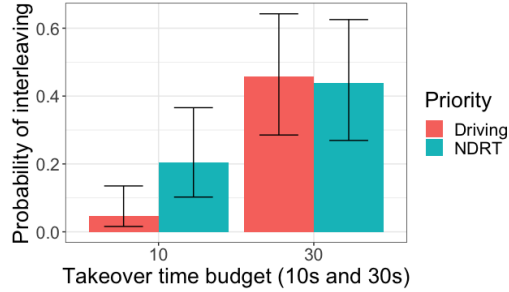

Fig. 4. Relationship between priority (driving and NDRT) and probability of adopting interleaving strategy during takeover moderated by takeover time budgets. Error bars present 95 Percent Confidence Limits.

*5.6.1 Effects of priority & takeover time-budget on takeover strategy (RQ2.1 & RQ2.2).* Both priority and takeover time budget were significant predictors of the probability of drivers interleaving between driving and NDRT during takeovers (Table 3). There was a significant interaction between priority and time budget. In other words, the effect of priority on the takeover strategy was moderated by the takeover time budget. In short (10s) time budget conditions, people were more likely to adopt an interleaving strategy when asked to prioritize NDRT. In long (30s) time budget conditions, the probability of interleaving was not affected by priority (Figure 4).

*5.6.2 Effects of takeover strategy on takeover performance & engagement in multi-step texting task (RQ2.3 & RQ2.4).* Similar to the first experiment, we did not find any significant relationship between takeover strategy and takeover quality (SDLP: $\chi 2(1) = 0.73, p = 0.39$ and SDV: $\chi 2(1) = 0.12, p = 0.73$). Neither of the control variables (priority and time budget) was significantly related to SDLP. However, prioritizing NDRT resulted in a larger SDV ($\chi 2(1) = 4.4, p = 0.03$). The takeover strategy had a significant effect on takeover timing, both for ADT ($\chi 2(1) = 4.25, p = 0.04$) and GRT ($\chi 2(1) = 57.01, p < 0.001$). Drivers took longer to resume driving even though they were quicker to glance at the driving scene after the takeover request when adopting an interleaving takeover strategy (Table 4). Among the control variables, time budget was a significant predictor in both models (ADT: $\chi 2(1) = 124.68, p < 0.001$, GRT: $\chi 2(1) = 54.15, p < 0.001$), but priority did not have a significant effect. NDRT engagement measures (number of questions asked: $\chi 2(1) = 0.004, p = 0.95$, correct guesses: $\chi 2(1) = 0.12, p = 0.73$) were not related to takeover strategies. Among control variables, only the time budget was a significant predictor for NDRT engagement (Number of questions asked: $\chi 2(1) = 15.85, p < 0.001$, Correct guesses: $\chi 2(1) = 5.56, p = 0.02$).

Table 4. Marginal means estimated from the corresponding models. Results are averaged over the two levels of priorities and two levels of takeover time budget.

|  |  | Suspension Mean(SE) | Interleaving Mean(SE) |  |
|---|---|---|---|---|
| Takeover quality | SDLP (cm) | 17(1.20) | 18(1.40) | t(216)=-0.85, p=0.39 |
|  | SDV (kmph) | 3.89(0.28) | 3.79(0.32) | t(216)=0.35, p=0.73 |
| Takeover timing | ADT (sec) | 11.80(0.98) | 13.30(1.09) | **t(215)=-2.06, p=0.04** |
|  | GRT (sec) | 10.20(0.96) | 3.4(1.11) | **t(219)=7.51, p<0.001** |
| NDRT engagement | Questions asked | 6.55(0.29) | 6.53(0.33) | t(241)=0.07, p=0.95 |
|  | Correct guess | 1.15(0.07) | 1.18(0.10) | t(249)=-0.34, p=0.73 |

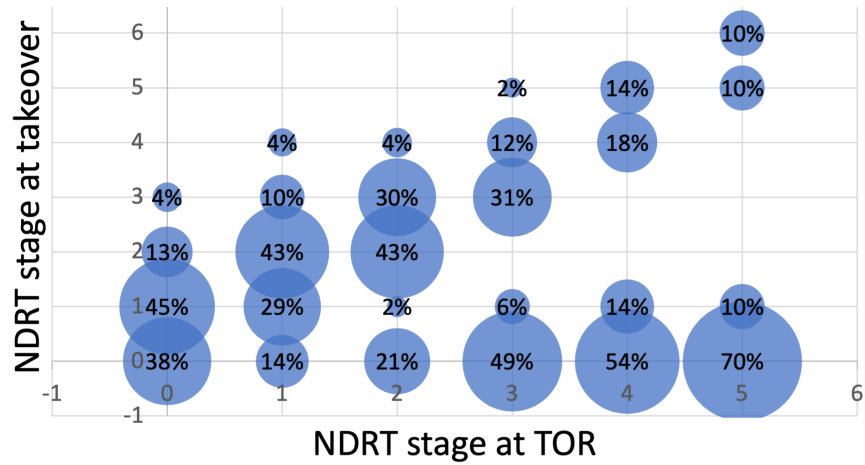

Fig. 5. NDRT stages at TOR and at takeover. The x-axis shows the stage of the NDRT when the TOR arrived. The y-axis shows the NDRT stage when driver took over manual driving. The circle-area shows the percentage of instances participants stopped at a certain NDRT for takeover while being at a certain NDRT stage during TOR.

*5.6.3 Effect of the stage of NDRT at TOR (RQ2.5).* Figure 5 shows the proportion of instances of various stages of NDRT at TOR where the participant stopped at a specific stage of NDRT during the takeover. We only considered NDRT stages that occurred at least 10 times among the participants at the time of TOR. We found that participants were more likely to complete the NDRT and then switch to driving if they were at later stages of the NDRT when the TOR was presented. For example, if participants were at stage 1 of NDRT when the TOR was presented, the proportion of time they switched to driving after finishing the NDRT was significantly lower compared to if they were at stages 3 ($\chi2$ (1)=11.41, p<0.001), 4 ($\chi2$ (1)=11.44, p<0.001), or 5 ($\chi2$ (1)=12.41, p<0.001) of NDRT at TOR. Similarly, if participants were at stage 2 of NDRT at TOR, they were significantly less likely to switch to driving after finishing the NDRT compared to if they were at stages 3 ($\chi2$ (1)=8.05, p=0.004), 4 ($\chi2$ (1)=8.23, p=0.004), or 5 ($\chi2$ (1)=9.35, p=0.002) of NDRT at TOR. If the participants were at stage 0 (finished current NDRT) at TOR, they were also more likely to switch to driving while being at stage 0 (not starting next NDRT) compared to if they were at stage 1 ($\chi2$ (1)=6.38, p=0.01) or 2 ($\chi2$ (1)=3.69, p=0.05). Taken together, this indicates that drivers tended to take over after finishing the NDRT they were performing if possible. This was more difficult to do when drivers were at the initial stages of the NDRT compared to later stages or stage 0 where they could just take over without starting a new NDRT.

## 6 DISCUSSION

Our results indicate that the cognitive demand of the NDRT does not influence the strategy the driver uses for taking over. However, this finding may change for a more complex and cognitively demanding driving scenario compared to the simple driving scenario we used in this experiment. Iqbal et al. found that the cognitive demand of NDRT affected driving performance only in complex driving scenarios [24]. So, the influence of the cognitive demand of NDRT on takeover strategy will also need to be investigated for more complicated driving scenarios.

Consistent with findings from a previous study by our team [45], our analyses show that people are more likely to interleave between driving and NDRT during takeovers if they are allowed a longer time for taking over. This finding was consistent for both texting conversation and multi-step texting tasks we used in the two experiments. Priority also affected the likelihood of interleaving during takeovers. When drivers were asked to prioritize NDRT, they were more likely to interleave compared to when they were prioritizing driving. Perhaps more interestingly, there was a significant interaction between priority and takeover time budget. The effect of priority on the likelihood of interleaving can only be observed for the 10-second takeover time budget but not for the 30-second condition. This suggests that in time-critical driving scenarios when drivers have to quickly take over driving, other factors like how they prioritize the NDRT can influence when they actually stop the NDRT and start driving.

The effects of takeover strategies were consistent in the two experiments we conducted. We did not see any significant difference in takeover quality in terms of SDLP and SDV between the suspension and interleaving strategies for takeovers. One reason for this could be that participants drove on a straight highway with no traffic on the road. In more complicated scenarios, where it takes more effort to control the vehicle after the takeover, the takeover strategy could influence takeover quality. This reasoning is also supported by our finding that the takeover strategy significantly affects takeover timing in terms of both ADT and GRT. When adopting an interleaving strategy, drivers took longer to start driving, but the first glance at the road after TOR was quicker compared to when they adopted a suspension strategy. If drivers decided to continue NDRT for a while before resuming driving, they would spend the extra time going back and forth between driving and NDRT. Such interleaving can improve drivers' situational awareness [15], which in turn may improve takeover quality in more challenging driving scenarios.

For the multi-step NDRT in the second experiment, we observed drivers' tendency to try and complete the NDRT before switching to driving. This is not surprising: prior research shows that people often switch to an interrupting task when they reach a boundary in an ongoing task, whether they are playing cards [75], or engaging in an NDRT in a car [27]. However, it can be dangerous if the driver takes too much time trying to finish the NDRT before taking over, especially since it is not always clear how long the NDRT will take to finish.

### 6.1 Limitations

While our study provides important insights into takeover strategy in conditionally automated vehicles, there are important limitations to mention. Our experiments were conducted using a (high-fidelity) driving simulator. Drivers' behavior in a vehicle on the road may differ from those we observed in our experiments. Nevertheless, similar to what is commonly observed in driving simulator studies [33], we noticed that our participants were actively engaged in the driving task. Another limitation is that the drivers engaged in both driving and NDRT for relatively short periods of time during the experiments. Their behavior may change if they drove or performed NDRTs for a longer period of time before switching, which will sometimes be the case in future conditionally automated vehicles. We also used a straight road without any traffic in our experiment to minimize the effects of other factors that were not of interest to this study.

So, the findings of this study will also have to be examined in future studies using more complex driving scenarios. Lastly, our experiments had a relatively small sample size and most participants were young college students. We also did not take into consideration how much driving experience participants had. So, it will be important to examine whether the findings of this study remain consistent for different groups of people under different contexts.

## 7 CONCLUSION

We expect that conditionally automated (SAE Level-3) vehicles will be available to consumers and that drivers will engage in non-driving-related tasks (NDRTs) when these vehicles are controlled by automation. Our work helps us understand how drivers will switch from the NDRTs to driving, once they receive a takeover request (TOR). First, for a variety of texting-based NDRTs, we found that drivers will often interleave between the NDRT and driving before they fully start manual driving. We also found that their decision to interleave or to stop the NDRT and turn directly to driving will be influenced by the time available to switch, but crucially also the priority assigned to the NDRT and driving, and the timing of when the TOR arrives with respect to the driver's progress in the NDRT. The time that is available to switch to driving will likely be set by the automation technology. However, which task has priority, and the timing of interruptions, as well as a host of other variables, can be affected by the tools that drivers will use for the NDRTs. Shaping these tools provides an opportunity for designers (and regulators) to support safe driving practices.

## ACKNOWLEDGMENTS

This work was funded in part by the National Science Foundation under grants CMMI-1840085 and CMMI-1840031.